\documentclass[twocolumn,twoside]{svmultivs_gm} 
\usepackage{graphicx}
\usepackage{hyperref}
\usepackage{comment}

\title*{The Wetterstein Millimeter Telescope: A New German Facility for Astronomy and Geodesy}
\titlerunning{WMT -- Neidhardt, Kadler et al.}
\author{Alexander Neidhardt$^{1}$, Matthias Kadler$^2$, A.K.-Baczko$^2,^4$, Guido Dietl$^2$, Urs Hugentobler$^1$, Matthias Schartner$^1$, Tobias Ullmann$^2$, Till Rehm$^3$, and Jompoj Wongphechauxsorn$^2$ for the WMT consortium}
\authorrunning{WMT -- Neidhardt, Kadler et. al.} 
\authoremails{alexander.neidhardt@tum.de, matthias.kadler@uni-wuerzburg.de}
\institute{$^1$FESG - Technische Universit{\"a}t M{\"u}nchen, $^2$Julius-Maximillians-Universit{\"a}t W{\"u}rzburg, $^3$ UFS Schneefernerhaus, $^4$~MPIfR Bonn}
\ContactAuthorName{Matthias Kadler}
\ContactAuthorTelephone{+49 931 31-85031}
\ContactAuthorEmail{matthias.kadler@uni-wuerzburg.de}
\NumberofInstitutions{2}
\InstitutionPostAddress{1}{Technische Universit{\"a}t M{\"u}nchen,
Forschungseinrichtung Satellitengeod{\"u}sie,
Geod{\"u}tisches Observatorium Wettzell,
Sackenrieder Straße 25
93444 Bad K{\"o}tzting}
\InstitutionCountry{1}{Germany}
\InstitutionWebPage{1}{https://www.asg.ed.tum.de/iapg/startseite/}
\InstitutionPostAddress{2}{Lehrstuhl für Astronomie,
Campus Hubland Nord,
Emil-Fischer-Straße 31,
97074 W{\"u}rzburg}
\InstitutionCountry{2}{Germany}
\InstitutionWebPage{2}{https://www.physik.uni-wuerzburg.de/astro/mitarbeiter/ag-kadler/prof-dr-matthias-kadler/}
\begin{document}  
\maketitle       
\abstract{The Wetterstein Millimeter Telescope (WMT) is a planned broadband (1.2--120 GHz) radio telescope to be established near the Environmental Research Station Schneefernerhaus (UFS) on Germany's highest mountain, the Zugspitze. Developed by a consortium of German research institutes and partners, the WMT is conceived as a multidisciplinary research infrastructure supporting radio astronomy, geodetic VLBI, satellite communications, space situational awareness, and technology development. The telescope is designed to operate within international VLBI networks, including the European VLBI Network, the Global mm-VLBI Array, and future ngVLA and SKA-VLBI observations. This contribution summarizes recent progress in the WMT project, including the evolution of the antenna design, and highlights the potential of the WMT to support future astronomical and geodetic VLBI.}
%
%
%
\section{Introduction}
%
\begin{figure*}[ht!]
    \renewcommand\thefigure{1}
    \begin{center}
         \includegraphics[width=0.9\textwidth]{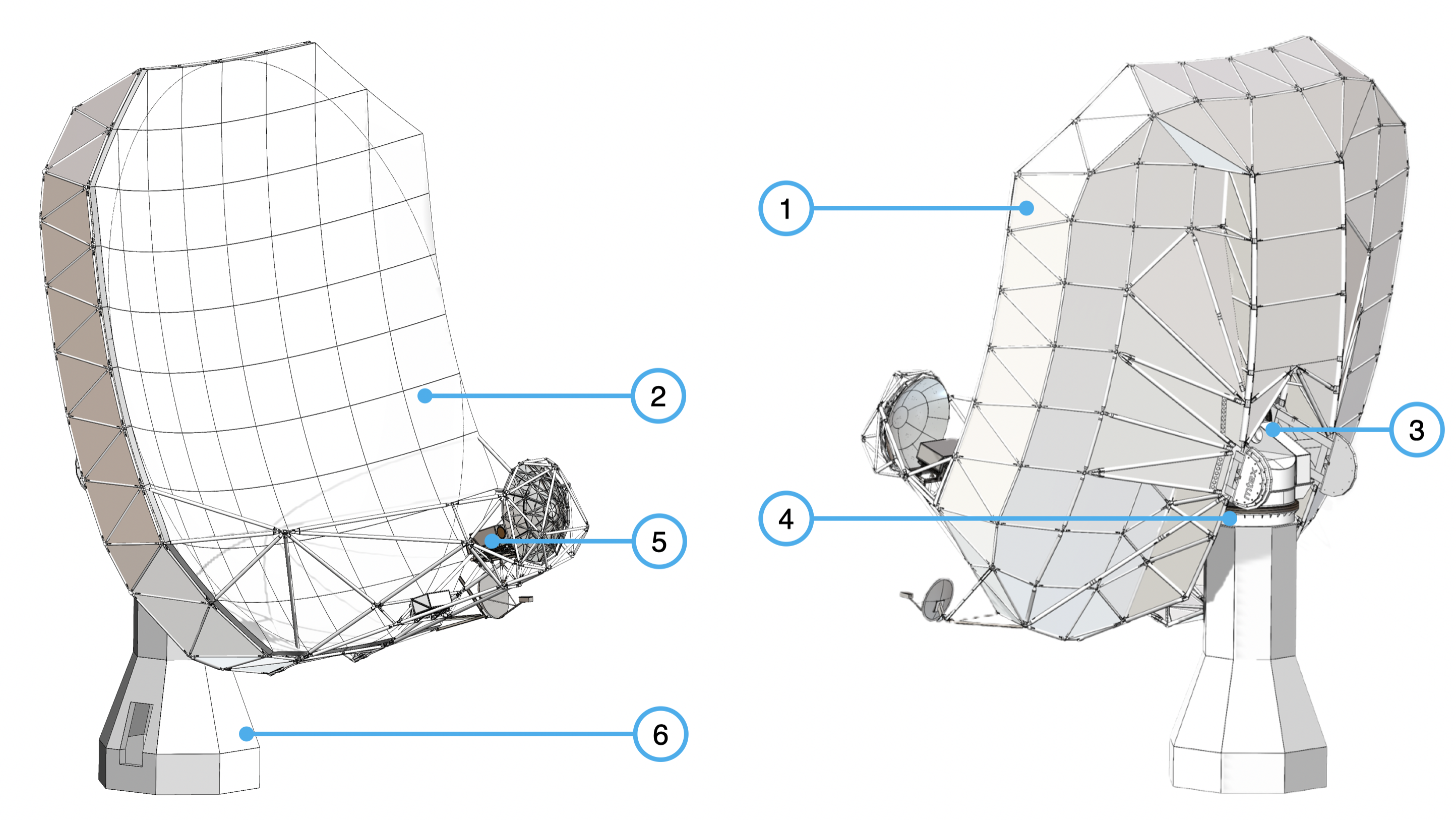}
    \end{center}
         \caption{Three-dimensional model of the current WMT baseline antenna design, highlighting the  modifications relative to the ngVLA prototype antenna: (1) reflector backside cladding of the backup structure, (2) main reflector panel heating, (3) segmented yoke steel structure, (4) azimuth bearing and drive support structure, (5) multi-functional front-end, and (6) concrete tower. Copyright: mtex antenna technology.}
         \label{fig:wmtdesign}
\end{figure*}

In recent years, Very Long Baseline Interferometry (VLBI) at centimeter and millimeter wavelengths has enabled groundbreaking discoveries in astronomy and forms an essential fundament for space geodesy. While radio astronomy has achieved the first images of the immediate surroundings of black holes, geodetic VLBI remains one of the fundamental space-geodetic techniques for realizing the terrestrial and celestial reference frames, determining Earth orientation parameters, and monitoring global geodynamics. The next generation of radio astronomy facilities will further extend these capabilities. The next-generation Very Large Array (ngVLA) is expected to provide unprecedented sensitivity at frequencies up to approximately 120\,GHz, while the Square Kilometre Array (SKA) will transform long- to mid-wavelength radio astronomy and enable highly sensitive SKA-VLBI observations.

The Wetterstein Millimeter Telescope (WMT) is being developed as a broadband (1.2–120 GHz), 18-m radio telescope on Germany's highest mountain, the Zugspitze, with the telescope site located at an altitude of 2,430\,m. The high-alpine site enables observations at millimeter wavelengths while providing close integration with the nearby Environmental Research Station Schneefernerhaus (UFS;  \cite{bittner2024}), creating unique opportunities for atmospheric monitoring, environmental measurements, and interdisciplinary research. The WMT is designed to operate both as a standalone observatory and as a station within existing and future VLBI networks.

The WMT has been conceived as a multidisciplinary research infrastructure. In addition to radio astronomy, the facility will support geodetic VLBI, satellite communications, and space situational awareness (SSA), while providing interfaces to future research infrastructures in astroparticle physics and gravitational-wave science. The observatory will also serve as a platform for the development of advanced instrumentation and next-generation VLBI techniques.

This paper presents the current status of the WMT project, with particular emphasis on the evolution of the antenna design and the opportunities provided by the observatory for geodetic VLBI.
\section{WMT Antenna Design}
The current baseline design of the WMT antenna is based on the 18-m offset Gregorian design of the ngVLA prototype antenna developed by mtex antenna technology GmbH for the ngVLA main array. While preserving the core optical and mechanical concept of the original design, several engineering modifications have been introduced as part of a feasibility study (conducted by mtex). This ensures that the requirements of year-round operation at the high-alpine WMT site is met, supporting the multidisciplinary scientific and technical applications foreseen for the facility, including radio astronomy, geodetic VLBI, satellite communications, and radar-based experiments. Beyond their importance for the WMT itself, these developments represent an engineering evolution of the original ngVLA antenna concept. They demonstrate engineering solutions for reliable operation under demanding alpine environmental conditions and provide valuable design experience for future high-performance ngVLA-type radio telescopes at climatically challenging sites. The principal design modifications of the current WMT baseline design are summarized in Fig.~\ref{fig:wmtdesign}.

To ensure reliable operation under severe winter conditions, the backup structure (BUS) is enclosed by a lightweight rear cladding that protects the reflector support structure from snow and ice accumulation. Although the BUS is designed to withstand the associated loads, accumulated snow and ice would degrade the antenna surface accuracy through increased structural deformation. The enclosed volume additionally accommodates a distributed heating and ventilation system that provides uniform heat distribution throughout the BUS and enables efficient de-icing of the reflector panels while minimizing thermal gradients. Together, the rear enclosure and the reflector panel heating system significantly improve the antenna's operational availability under alpine winter conditions while preserving the reflector surface accuracy.

Several major structural components were redesigned to satisfy the strict transport constraints imposed by the mountain site, limiting components to a maximum mass of approximately 4--5\,t. The welded yoke structure was therefore segmented into transportable modules, while its floor was redesigned from a solid steel plate to a lightweight beam-and-grid construction, reducing the structural mass from 15.3\,t to 12.9\,t. Likewise, the original ngVLA upper pedestal carrying the azimuth bearing and drive system was replaced by a compact azimuth bearing and drive support structure only 600\,mm high and weighing less than 4\,t. This assembly is mounted directly onto a reinforced-concrete tower via an anchor ring derived from the original ngVLA foundation concept. Replacing the steel pedestal with a concrete tower considerably simplifies transportation, on-site assembly, structural alignment, and thermal insulation, making it better suited to construction and long-term operation under high-alpine conditions.

Beyond these site-driven adaptations, the current WMT baseline design also incorporate a multi-functional front-end. It is designed to accommodate not only astronomical receivers but also instrumentation supporting additional applications This includes geodetic instrumentation, radar experiments, satellite communication technologies and other multidisciplinary use cases envisioned for the WMT.
The resulting flexibility also facilitates the integration of instrumentation required for future geodetic VLBI developments and other technical demonstrations without compromising the primary astronomical capabilities of the observatory.

Taken together, these modifications preserve the proven performance of the ngVLA antenna concept while extending it for reliable operation in a demanding alpine environment and broadening its functionality beyond a dedicated radio astronomy instrument. The resulting baseline design provides a versatile platform for high-frequency radio observatories operating under climatically demanding environmental conditions while accommodating the broader range of scientific and technical applications envisioned for the WMT.

\section{Science goals}
The WMT is being developed as a multidisciplinary research infrastructure combining radio astronomy, astronomical and geodetic VLBI, high-frequency engineering, and experimental applications in radar and satellite communications. Its broadband frequency coverage from approximately 1.2 to 120\,GHz, its flexible instrumentation concept, and its high-alpine location allow a single facility to support fundamental research, applied science, and the development and validation of new observational techniques.

The scientific and technological objectives of the WMT can be grouped into three principal areas:

\begin{itemize}
\item \textbf{Radio astronomy:} millimeter-wavelength variability and VLBI studies of relativistic jets in active galactic nuclei, electromagnetic follow-up of multi-messenger signals and transient radio sources at high radio frequencies, galaxy evolution and cosmic structure formation, as well as solar physics and space-weather studies;

\item \textbf{Geodesy and Earth-related applications:} support of geodetic VLBI to realize and improve terrestrial and celestial reference frames, investigations of source structure and atmospheric propagation effects, and the development of observing and calibration methods relevant to precision geodesy;

\item \textbf{Technology and applied research:} satellite communications and high-frequency engineering, multistatic radar techniques for space situational awareness, monitoring of radio frequency bands to protect against unwanted emissions, e.g., from global satellite constellations, and the development and validation of technologies for future radio-astronomical and geodetic infrastructures.
\end{itemize}

The combination of these objectives is a driver for the whole WMT project. Rather than being optimized exclusively for a single scientific discipline, the facility is conceived from the outset as a shared platform for astronomy, geodesy, engineering, and applied research. Its broadband receiver concept and multifunctional front-end architecture allow the integration of different instruments and observing modes to be integrated without restricting the primary radio-astronomical capabilities of the telescope. This approach maximizes the scientific return of the facility and fosters methodological and technological exchange between communities that have traditionally operated largely separate infrastructures.

The high-altitude location on Zugspitze and the proximity of the UFS provide unique opportunities for atmospheric studies and calibration methods relevant to both astronomical and geodetic VLBI.

Beyond the initial WMT facility, the project may also provide technological and operational experience for the longer-term LEVERAGE (Long-baseline Extension in next-generation VLBI Experiments and Rapid-response Array Germany) concept. LEVERAGE envisions a distributed network of mid- to high-frequency antenna clusters in Germany, with possible future extensions across Europe. Such a network could complement existing and future astronomical and geodetic VLBI infrastructures and improve the long-baseline $(u,v)$ coverage of the ngVLA Long Baseline Array, particularly on baselines exceeding 5000\,km \cite{LEVERAGE}.

\section{The WMT in the International VLBI Landscape}
The WMT is designed to operate both as a stand-alone single-dish station and as a station in existing and future VLBI networks. Owing to its geographic location in southern Germany, broadband frequency coverage, and high-frequency capabilities, the telescope provides a valuable extension to several international VLBI infrastructures, including
\begin{itemize}
\item the European VLBI Network (EVN),
\item the Global Millimeter VLBI Array (GMVA),
\item the Event Horizon Telescope (EHT),
\item the ngVLA, 
\item future Square Kilometre Array (SKA) VLBI, and
\item the IVS especially in the K-band or for source structure analysis (see below)).
\end{itemize}

The long baselines between the WMT and existing European, North American, and African stations substantially improve the $(u,v)$ coverage of these networks, particularly at high observing frequencies. This leads to improved imaging fidelity and increased angular resolution while strengthening the European contribution to global VLBI observations.

Beyond its role within international arrays, the WMT also opens the possibility of establishing a dedicated German high-frequency VLBI capability. Such a facility would provide increased flexibility for technology demonstrations, commissioning activities, rapid-response observations, and the development of new astronomical and geodetic VLBI techniques.

\section{Relevance for geodetic VLBI}
The WMT shares several characteristics with modern VGOS antennas, most notably its broadband receiving concept that enables its application to a broad range of geodetic VLBI observations. It can support legacy S/X, as well as VGOS and K-band observations. The telescope is primarily designed for astronomical observations, and therefore does not achieve the slewing rates of a dedicated VGOS system. The space requirements for heavy cryogenic receivers in the prime focus of the telescope, together with other astronomical receivers, might be challenging. For this reason, the WMT is not intended to replace dedicated VGOS antennas in routine IVS operations. Instead, it complements the existing geodetic infrastructure by providing broadband high-frequency observations, source-structure monitoring, technology development, and unique opportunities arising from its co-location with atmospheric and environmental instrumentation.
Co-located GNSS receivers, nearby atmospheric sensors, and access to high-resolution weather model data from the adjacent UFS can all contribute to improving the quality and interpretation of  observations. 

The WMT can make a major impact by bridging geodetic and astronomical VLBI and taking on tasks in the areas of astrometry in general, such as source structure monitoring, and source variability studies. An important factor in improving observation models and scheduling is a deeper understanding and continuous monitoring of the intrinsic dynamics of the extragalactic radio sources themselves \cite{sourcestructure}. Although geodetic VLBI primarily observes carefully selected active galactic nuclei (quasars), many of these sources exhibit complex and temporally variable jet structures that vary over time. Their frequency-dependent morphology gives rise to astrometric position shifts and additional group-delay errors. Recent studies have shown that source structure effects can already affect geodetic VLBI observations at the level of several millimetres, thereby representing a limiting factor for future ultra-precise terrestrial and celestial reference frame realizations. 

Investigations of core-shift effects, jet collimation, and the temporal evolution of relativistic jets not only provide important insights into the physics of relativistic outflows but also deliver essential input for the continued refinement of geodetic reference-source models. In particular, high-angular-resolution observations of compact AGN jets enable a precise characterization of frequency-dependent core positions (core shift), temporal structural variations, and their impact on astrometric and geodetic VLBI observables. In this context, the WMT can make a significant and visible contribution to research projects on source structure and its temporal evolution. These contributions and analyses are crucial for achieving the precision envisioned for future geodetic VLBI applications. A better understanding of the physics in and around active galactic nuclei provides information directly applicable to geodetic reference-source modelling that can be supported by the WMT. In addition, it fits ideally into the IVS-supported K-band observations of the ICRF.

Finally, the telescope infrastructure can serve as a testbed for novel receiver, front-end, and back-end technologies. This includes front-end sampling architectures, their calibration and characterization, as well as advanced time and frequency metrology. High-precision time and frequency references, together with their distribution systems, are expected to play an increasingly important role in future astronomical and geodetic applications. Accordingly, these technologies will also constitute an integral part of the research activities within the WMT project. This technically oriented scientific focus enables projects aimed at improving future geodetic facilities and geodetic product pipelines.

\section{Conclusions and outlook}
The Wetterstein Millimeter Telescope (WMT) is being developed as a multidisciplinary high-frequency radio observatory that combines radio astronomy, geodetic VLBI, technology development, and applied research within a single research infrastructure. The current baseline design represents a significant evolution of the original ngVLA antenna concept, addressing the demanding environmental conditions of the high-alpine site while preserving the performance required for high-frequency observations.

Beyond its astronomical science programme, the WMT will provide valuable opportunities for geodetic VLBI through source structure monitoring, technology development, and its unique co-location with atmospheric and environmental instrumentation. Together with its broadband capabilities and flexible instrumentation concept, these features establish the WMT as a complementary facility for existing and future international VLBI networks.

Since construction work is limited to the period between late spring and fall each year, the following construction plan is currently foreseen:
\begin{itemize}
\item 2027: site preparation, civil engineering, and foundation construction;
\item 2028--2029: antenna assembly, installation, and start of commissioning;
\item from 2030 onward: early science operations, initial participation in international VLBI observations, and progressive integration into next-generation VLBI infrastructures.
\end{itemize}
The WMT is intended to become a long-term platform for scientific discovery, technological innovation, and interdisciplinary collaboration, strengthening both the astronomical and geodetic VLBI communities in Germany and internationally.
%
\par\medskip
\textbf{Acknowledgements: }
{\small The authors gratefully acknowledge the contributions of mtex antenna technology GmbH to the development of the WMT antenna design. The WMT project is further supported by the 2026 parliamentary initiative (Fraktionsinitiative) of the Bavarian State Parliament (Bayerischer Landtag), which enabled key preparatory activities for the project.
The authors gratefully acknowledge the invitation to present the WMT project at the 14th General Meeting of the International VLBI Service for Geodesy and Astrometry (IVS GM2026) in Garmisch-Partenkirchen.}

%
%
%
%

%
\end{document}